\documentstyle[12pt]{article}
\begin{document}
\baselineskip 22pt plus 2pt
\begin{center}
{\bf ORBITAL MAGNETISM IN TWO-DIMENSIONAL INTEGRABLE SYSTEMS
\vspace{1.5cm}

E. Gurevich and B. Shapiro} \\

Department of Physics \\
Technion-Israel Institute of Technology \\
Haifa 32000, Israel
\vspace{1.5cm}

{\bf Abstract} \\
\end{center}

We study orbital magnetism of a degenerate electron gas in a number
of two-dimensional integrable systems, within linear response theory.
There are three relevant energy scales: typical level spacing
$\Delta$,
the energy
$\Gamma$,
related to the inverse time of flight across the system, and the
Fermi energy
$\varepsilon_F$.
Correspondingly, there are three distinct temperature regimes:
microscopic
$(T \ll  \Delta)$,
mesoscopic
$(\Delta \ll T$ \raisebox{-.6ex}{\mbox{$\stackrel{\textstyle <}{ \sim}$}} $\Gamma )$
and macroscopic
$(\Gamma \ll T \ll \varepsilon_F)$.
In the first two regimes there are large finite-size effects in  the
magnetic susceptibility
$\chi$,
whereas in the third regime
$\chi$
 approaches its macroscopic value. In some
cases, such as a quasi-one-dimensional strip or a harmonic confining
potential, it is possible to obtain analytic expressions for
$\chi$
in the entire temperature range.
\pagebreak

\noindent {\bf 1. \ Introduction} \\

A degenerate electron gas, in the presence of a weak magnetic field,
exhibits weak orbital magnetism$^1$ (the Landau diamagnetism). For a
two-dimensional gas the value of the orbital magnetic susceptibility
is given by
$\chi_L = - e^2/12 \pi Mc^2$,
where $M$ is the electron mass. (Double degeneracy with respect to spin
is implied in this expression, as well as in all subsequent
formulae.) This result applies to a {\it macroscopic} system. 

Whether a sample of a given size $L$ can be considered as macroscopic,
depends on the temperature $T$. Namely, $T$ (in energy units) should be
compared with the characteristic size-dependent energies such as the
mean level spacing,
$\Delta \simeq 2 \pi \hbar^2/ML^2$,
or the inverse time of flight across the sample,
$\Gamma \simeq \hbar v_F/L \simeq k_F L \Delta /2 \pi$,
where
$v_F$
and
$k_F$
are the Fermi velocity and wave number. Generally, one should
distinguish between three temperature regimes: 

\begin{enumerate}
\item[(i)] \ For
$T < \Delta$
(the ``microscopic'' regime) discreteness of the energy levels comes
into play and the sample can be viewed as a giant atom. The magnetic
response in this case can be very strong and includes such exotic
possibilities as perfect diamagnetism and Meissner effect$^2$. 

\item[(ii)] \ For higher temperatures,
$\Delta \ll T$ \raisebox{-.6ex}{\mbox{$\stackrel{\textstyle <}{ \sim}$}}
 $\Gamma$,
the system enters the mesoscopic regime (for recent reviews see
Refs 3,4). Here the typical value of the magnetic susceptibility
is of order
$(k_FL)^{\alpha} \mid \chi_L \mid$
and can have either sign. The exponent
$\alpha$
depends on the sample geometry. For most two-dimensional integrable
systems
$\alpha = 3/2$,
although other values are also possible in some special cases (see
below). For completely chaotic two-dimensional systems
$\alpha = 1$. 

\item[(iii)] \ For still higher temperatures, when
$T \gg \Gamma$
(but smaller than the Fermi energy
$\varepsilon_F)$,
the system can be considered as macroscopic and its magnetic
susceptibility, up to small corrections, is given by the Landau value
$\chi_L$. 
\end{enumerate}

Thus, at present there is a good qualitative understanding of the
phenomenon of orbital magnetism in various temperature regimes and for
various geometries. However, reliable quantitative results are scarce.
Most of such results refer to the mesoscopic regime$^{4-6}$ and are
based on a semiclassical approximation for the density of states. This
approximation becomes inadequate both at very low temperatures,
$T < \Delta$,
and at high temperatures,
$T > \Gamma$.
It, therefore, seems useful to consider a few simple systems, starting
with an exact expression for the susceptibility
$\chi$,
and to observe the behaviour of
$\chi$
in the entire temperature range. This is done in the present paper
by using a linear response expression for
$\chi$. 

In Section 2 we present several equivalent expressions for the orbital
magnetic susceptibility within the linear response theory. In Sections
3, 4, 5 we consider specific examples of a strip, disc, and
square-geometries. Section 6 is devoted to an electron gas confined
by a two-dimensional parabolic potential.
\vspace{0.5cm}

\pagebreak

\noindent {\bf 2. \ Linear Response Theory for the Magnetic
Susceptibility} \\

We consider an electron gas, confined to some domain in the xy-plane and
subjected to a weak magnetic field $B$ in the z-direction. The
grand-canonical potential
\begin{eqnarray}
\Omega = -
\frac{1}{\beta}
\int dE \rho (E) \ln
[1 + e^{\beta (\mu-E)} ] \ ,    
\end{eqnarray}
where
$\mu$
is the chemical potential,
$\beta = 1/T$
and
$\rho (E)$
is the single-particle density of states. Sometimes it is useful,
by integrating by parts twice, to rewrite Eq. (1) as:
\begin{eqnarray}
\Omega = -
\int dE \Omega (E)
\frac{\partial f}{\partial E}     
\end{eqnarray}
where
$f(E) = \{ 1 + \exp [\beta (E - \mu)] \}^{-1}$
is the Fermi function and the quantity
\begin{eqnarray}
\Omega (E)
= - \int^E_{- \infty}
dE^{\prime}
\int^{E^\prime}_{- \infty}
dE^{\prime\prime}
\rho (E^{\prime\prime})      
\end{eqnarray}
has the meaning of a grand-canonical potential, for the same system,
at zero temperature and with the chemical potential equal to $E$. 

The density of states can be written as
\begin{eqnarray}
\rho (E) = -
\frac{1}{\pi}
Im Tr G(E)
= -
\frac{1}{\pi}
Im Tr
\left(E + i\eta - H\right)^{-1} \ ,      
\end{eqnarray}
where
$G(E)$
is the retarded Green's function, at energy $E$. The full
(single-particle) Hamiltonian $H$ is split into the unperturbed part,
$H_o = \hat{p}^2/2M$,
and the perturbation
\begin{eqnarray}
V  = -
\frac{e}{Mc}
\vec{A} \cdot \vec{p}
+
\frac{e^2}{2Mc^2} A^2 \ ,      
\end{eqnarray}
which describes the effect of the (static) magnetic field. It is assumed
that the vector potential
$\vec{A}$
satisfies the condition
$\mbox{div} \vec{A} = 0$. 

Expanding
$G(E)$
in terms of the unperturbed Green's functions,
$G_o(E) = (E + i \eta - H_o)^{-1}$,
one obtains, up to second order,
\begin{eqnarray}
G = G_o + G_o V G_o + G_o V G_o V G_o \ ,    
\end{eqnarray}
which, on substitution into Eq. (4), leads to the following expression
for the correction
$\delta \rho (E)$
to the density of states:
\begin{eqnarray}
\delta \rho = -
\frac{1}{2 \pi}
\left( \frac{e}{Mc} \right)^2
\frac{\partial}{\partial E}
Im Tr \{ [ G_o
(\vec{A} \cdot \vec{p}) ]^2
+ M (G_o A^2) \} \ .          
\end{eqnarray}
The first and the second term describe, respectively, the para- and
diamagnetic corrections. Plugging Eq. (7) into (1) and integrating
by parts gives:
\begin{eqnarray}
\delta \Omega
= -
\frac{1}{2 \pi}
\left( \frac{e}{Mc} \right)^2
\int dE f(E) 
Im Tr \left\{ \left[ G_o
\left(\vec{A} \cdot \vec{p}\, \right) \right]^2 +  M\left(G_o A^2\right) 
\right\} \ .       
\end{eqnarray}
This expression is proportional to
$B^2$,
so that the susceptibility, in the
$B \rightarrow 0$
limit, is
$\chi = - 2 \delta \Omega / B^2 A$,
where $A$ is the sample area. 

For further calculations it is useful to have an expression for
$\chi$
in terms of
$\chi_o (E)$
which is the susceptibility of the system at T=0,
$\mu = E$
(compare with Eq. (2)):
\begin{eqnarray}
\chi = - \int dE \chi_o(E)
\frac{\partial f}{\partial E}     
\end{eqnarray}

Eq. (8) is written in an abstract operator form.  For practical
calculations it is useful to use a particular representation.  For
example, if one computes the traces using as a basis the
eigenstates of
$H_o$
(with the appropriate boundary conditions), one obtains, after some
algebra:
\begin{eqnarray}
\chi = -
\frac{1}{AB^2}
\sum_i
\left\{ 2f(\varepsilon_i) \varepsilon^{\prime\prime}_i +
\frac{\partial f}{\partial \varepsilon_i}
(\varepsilon^{\prime}_i)^2 \right\} \ .     
\end{eqnarray}
Here the summation is over all states of the unperturbed Hamiltonian
$H_o$, with eigenenergies
$\varepsilon_i$.
The energies
$\varepsilon^{\prime}_i$
and
$\varepsilon^{\prime\prime}_i$
denote the first (i.e. proportional to $B$) and the second
(proportional to $B^2$) corrections to the unperturbed energies
$\varepsilon_i$.
The first correction can exist only for levels which are degenerate
in the absence of $B$. 

Another useful expression for
$\chi$
is obtained by writing Eq. (8) in the coordinate representation.
This results in:
\begin{eqnarray}
\chi &=& 
\frac{1}{\pi AB^2}
\left( \frac{e}{Mc} \right)^2 
Im \int dE f(E) \int d^dr \int d^dr^{\prime} \times 
\mbox{\rule[-.8cm]{0cm}{1cm} } \nonumber \\ 
&& \times \left\{ - \hbar^2
\left[\vec{A}(\vec{r}) \cdot
\frac{\partial}{\partial \vec{r}}
G_o
(\vec{r}, \vec{r}^{\hspace*{0.05cm}\prime}; E)\right]
\left[\vec{A}(\vec{r}^{\hspace*{0.05cm}\prime}) \cdot
\frac{\partial}{\partial \vec{r}^{\prime}}
G_o
(\vec{r}^{\hspace*{0.05cm}\prime}, \vec{r}; E\right]\right. + 
\mbox{\rule[-.8cm]{0cm}{1cm} } \nonumber \\
&& \hspace{1cm} + M \left. \delta
(\vec{r} - \vec{r}^{\hspace*{0.05cm}\prime}) G_0(\vec{r},\vec{r}^{\hspace*{0.05cm}\prime};E)
A^2(\vec{r}{\hspace*{0.05cm}})\mbox{\rule[-.3cm]{0cm}{0.01cm} }\right\} \ ,    
\end{eqnarray}
where $G_o
(\vec{r}, \vec{r}^\prime; E)$ is the unperturbed (retarded) Green's function
 in the coordinate representation.  Since this representation is usually
 the most appropriate for making various approximations, Eq.~(11) is a good 
starting point
in many cases.  It was used, for instance, in the study of mesoscopic
 effects in 
disordered systems$^7$ (in this case $G_0$ includes the random potential of 
impurities).  It also helps to prove in the most direct way that, for 
$T>\Gamma$, the susceptibility approaches its macroscopic value $\chi_L$, 
independently of the sample geometry.  Indeed, the Fermi function $f(E)$
has poles at values $E_n=\mu+i(\pi/\beta)(2n+1)$.
Therefore the integral over $E$ in Eq.~(11) can be replaced by a sum
 containing
$G_0
(\vec{r}, \vec{r}^{\hspace*{0.05cm}\prime}; E_n)$.  This function in an
 infinite system decays with
distance as $\exp(-k_F\left|\vec{r}-\vec{r}^{\hspace*{0.05cm}\prime}
\right|\pi(2n+1)/\beta\mu)$.  
It is therefore clear that for a system with size $L$ larger than
$\beta\mu/k_F$, i.e. for $T>\Gamma$, the susceptibility ceases to 
depend on sample
size or on its geometry.  Therefore, for $T>\Gamma$ linear response
 theory is applicable
as long as the cyclotron energy $\hbar \omega_c$ is smaller than $T$. 
 However, for 
$T<\Gamma$ the susceptibility $\chi$ does depend on sample size and its
geometry, and the linear response condition requires that $\hbar \omega_c$ is 
smaller 
than the level spacing $\Delta$ (i.e. the magnetic flux $\Phi$ through
 the sample
is smaller than the flux quantum $\Phi_0=2\pi\hbar c/e)$.

A useful approximate expression for $\chi$ is obtained upon using the 
semi-classical approximation$^8$ for the Green's functions in Eq.~(11).  Let
us briefly outline the derivation (details are presented elsewhere$^9$).
First, one derives a semi-classical approximation for the function $\chi_0(E)$.
This is done by rewriting the Green's functions in terms of the propagators
$K(\vec{r},\vec{r}^{\hspace*{0.05cm}\prime},t)$, approximating the propagators by their
semi-classical expressions$^8$ and performing the integrals within a saddle-point
approximation.  Then one substitutes $\chi_0(E)$ into Eq.~(9) and integrates
over $E$, making use of the approximation
\begin{eqnarray}
-\int dE \, E^\alpha \cos (F(E)) \frac{\partial f}{\partial E} \approx
\mu^\alpha \cos \left(F(\mu)\right) 
{\cal R}\left(\pi F^\prime(\mu)T \right) \ , 
\end{eqnarray}
where $F(E)$ is some smooth function of $E$ (i.e. does not change appreciably
within an interval of order $T$) and ${\cal R}(x)\equiv x/\sinh x$.  The
resulting semi-classical expression for $\chi$ is 
\begin{eqnarray}
\frac{\chi}{|\chi_L|} = 24\pi AM\sum_{\lambda,r} \frac{1}{\tau^2_\lambda}
\frac{\langle A^2_\lambda\rangle}{A^2} d_{\lambda,r}(\mu) 
{\cal R}\left(\frac{r\tau_\lambda}{\tau_t} \right) \ .   
\end{eqnarray}
Here $\lambda$ labels families of primitive periodic orbits for  integrable
systems or isolated orbits for chaotic ones. $r$ is the winding number,
$\tau_\lambda$ is the period of a primitive periodic orbit and 
$\tau_t=\hbar/\pi T$.  Factors $d_{\lambda, r}(\mu)$ are related to the 
oscillating part of the (unperturbed) semi-classical density of states$^{8,10}$:
\begin{eqnarray}
\rho_{\rm osc}(\mu) = \sum_{\lambda,r} d_{\lambda,r}(\mu) \ .   
\end{eqnarray}
For integrable system $\langle A^2_\lambda\rangle$ is an orbit area squared and
averaged over the family $\lambda$.  For a chaotic system it is simply
the squared area of an isolated orbit.  Averaging over a family amounts
to integration over one of the angle variables$^4$
\begin{eqnarray}
\langle A^2_\lambda\rangle= \frac{1}{2\pi}\int^{2\pi}_0 A^2(\theta)d\theta \ . 
\end{eqnarray}
The semi-classical expression for $\chi$, Eq.~(13), coincides with the one
derived in Ref.~4, in the $B\rightarrow 0$ limit.  This fact demonstrates that
it does not matter which of the two approximations, i.e. linear response or
semi-classics, is done first.\\

\noindent{\bf 3. Strip Geometry}

In this section we consider electrons confined to a strip
$-L_x/2<x<L_x/2$,
$-L_y/2<y<L_y/2$.  
Periodic and zero boundary conditions are assumed along $x$ and $y$ 
directions respectively, and the limit $L_x \rightarrow \infty$ is taken.  
It is convenient to choose the Landau gauge:
$A_x=-B y$.  $A_y=A_z=0$.  Stationary states are labelled by two quantum
numbers, $-\infty < k < + \infty$ and $n=1,2,\dots$. The eigenfunctions
$\psi_{n,k}(x,y)= e^{ikx} u_{n,k}(y)$, where $u_{n,k}$ satisfies:
\begin{eqnarray}
\left[ - \frac{\hbar^2}{2M} \frac{\partial ^2}{\partial y^2} + \frac{1}{2M}
(\hbar k + \frac{e}{c} B y)^2\right]u_{n,k}(y) = \epsilon u_{n,k}(y) \nonumber
\end{eqnarray}
Treating the magnetic field as a  perturbation, one obtains$^{11}$ that the first
order correction $\epsilon_{nk}' = 0$ and the second order correction
\begin{eqnarray}
\epsilon^{\prime\prime}_{n k} = \frac{L^2_y}{24M}
\left(\frac{eB}{c}\right)^2\left[1-\frac{6}{\pi^2n^2}+\frac{k^2L_y^2}{\pi^4}
\left(\frac{\pi^2}{n^2}-\frac{15}{n^4}\right)\right] \ . 
\end{eqnarray}

Thus, only the first term in Eq.~(10) is present, and the zero-temperature
susceptibility (including spin degeneracy) $\chi_0(E)$ is given by
\begin{eqnarray}
\chi_0(E) = - \frac{4}{B^2L_y} \int^{+\infty}_{-\infty}
\frac{dk}{2\pi} \sum^\infty_{n=1}  \epsilon^{\prime\prime}_{n k}\theta(E-\epsilon_{nk}) \ , 
\end{eqnarray}
where $\epsilon_{n k} = (\hbar^2k^2/2M) + (\hbar^2\pi^2n^2/2ML^2_y)$ are
the unperturbed energy levels.  Next, we integrate over $k$, apply the Poisson
summation formula$^1$ to the sum over $n$, and insert the resulting expression for
$\chi_0(E)$ into Eq.~(9).  The integral over $E$ is then performed, using
the approximation (12), which amounts to neglecting small terms of order
$T/\mu$.  The final expression for $\chi$ is:
\begin{eqnarray}
\frac{\chi}{\chi_L} &=& 1+\sqrt{\frac{k_FL_y}{\pi} } \sum^\infty_{\ell=1}
\frac{\cos \left(2\ell k_F L_y - \frac{3\pi}{4}\right)}{\mbox{\rule[.1cm]{0cm}{0.3cm} }\ell^{{3}/{2}}}{\cal R}\left(\frac{2\pi \ell T}{\Gamma}\right) +
\mbox{\rule[-.6cm]{0cm}{1cm} } \nonumber \\
&& \hspace{6cm} +  \mbox{\rule[-.1cm]{0cm}{0.8cm} } O(1/\sqrt{k_FL_y}) \ , 
\end{eqnarray}
where $k_F=\sqrt{2M\mu/\hbar^2}$, $\Gamma=\hbar^2k_F/ML_y=\hbar v_F/L_y$ and the
function ${\cal R}(x)$ has been defined above.
In Eq.~(18) we wrote down only the leading oscillating term, of order
$\sqrt{k_FL_y}$, and the term 1, responsible for the Landau diamagnetism.
There exist also small oscillating corrections, of order $(k_FL_y)^{-{1}/{2}}$
and smaller which are not written down explicitly, even though they are calculable.$^9$
Let us only mention that, in addition to oscillating corrections, there is also
a small non-oscillating paramagnetic correction, $9 |\chi_L|/8 k_FL_y$, to
the Landau value $\chi_L$.

Thus, the strip geometry provides a rare example for which it is possible to
obtain an essentially exact (up to small corrections of order $T/\mu$) expression for the
linear susceptibility $\chi$, including all size-dependent terms.  One can observe
the change of $\chi$ with $T$ in the entire temperature range, from $T=0$
and up to $T \gg \Gamma$ when $\chi$ becomes equal to its macroscopic value
$\chi_L$.  Eq.~(18) is quite similar to the corresponding expression for the
case of a parabolic confinement.$^{12}$  This fact demonstrates that the nature of
confinement (i.e. hard walls or ``soft'' confinement) is immaterial for the phenomenon
of size-dependent fluctuations.

In fact, the leading oscillating term in $\chi$ can be obtained, within a semi-classical
approximation, for any confining potential and for arbitrary magnetic field.$^9$
For small fields, the oscillations are given by an expression similar to Eq.~(18),
up to an overall factor of order unity and an extra phase in the 
argument of the cosine.  $L_y$ should be understood as some effective width of the
confining potential.

\pagebreak

\noindent{\bf 4. Circular Geometry}

The electron gas is confined to a disc of radius $R$.  The unperturbed, i.e.
zero-field stationary states are given, up to a normalisation factor, by 
$\exp(im\theta) J_m(\lambda_{mn}r/R)$, where $\lambda_{mn}$ is the $n$-th zero
of a Bessel function $J_m(x)$.  The unperturbed energies are 
$\epsilon_{mn} = (\hbar^2/2MR^2)\lambda^2_{mn}$.  A pair of states $|m,n\rangle$ and
$|-m,n \rangle$ have the same energy.

The perturbation term in the Hamiltonian, due to the magnetic field, is:
\begin{eqnarray}
V = \frac{i\hbar eB}{2Mc} \frac{\partial}{\partial \theta} + \frac{e^2B^2}{8Mc^2}r^2 \ , 
\end{eqnarray}
and the first and second-order energy corrections are:
\begin{eqnarray}
\epsilon'_{mn} = - \frac{e\hbar B}{2Mc}m \ , \ \ \epsilon^{\prime\prime}_{mn}
= \frac{e^2B^2}{8Mc^2}\langle m n|r^2|m n \rangle \ . 
\end{eqnarray}
Note that the first-order term in $V$ does not contribute to the correction 
$\epsilon^{\prime\prime}_{mn}$ which is therefore purely diamagnetic. 
 It now follows from
Eq.~(10) that
\begin{eqnarray}
\frac{\chi}{|\chi_L|} = - \frac{6}{R^2}\left\{\sum_{mn}
 \langle mn|r^2|mn\rangle f(\epsilon_{mn}) + \frac{\hbar^2}{M} 
\sum_{mn} m^2 \frac{\partial f}{\partial \epsilon_{mn}}\right\} \ . 
\end{eqnarray}

We analyse first the low-temperature regime,
 $T\ll\Delta \equiv 2\hbar^2/MR^2$.
Since $\langle mn|r^2|mn\rangle \simeq R^2$, the first (diamagnetic) term is
of order of the total number of electrons, $N\simeq (k_FR)^2=4\mu/\Delta$.
The second (paramagnetic) term exhibits sharp peaks every time when the 
chemical potential $\mu$ coincides with an energy level $\epsilon_{mn}$.  
Indeed, for $\mu=\epsilon_{mn}$, the function
 $\partial f/\partial \epsilon_{mn}$ is equal
to $-1/4T$, and it rapidly decreases when $\mu$ deviates from 
$\epsilon_{mn}$ by a 
few $T$.  Since the quantum number $m$, for states near $\mu$, is typically
 of order $k_FR$, the height of the paramagnetic peaks is of order
 $(k_FR)^2\Delta/T
\simeq \mu/T$.  Thus, in the low-temperature regime the susceptibility $\chi$
(normalised to the Landau value $|\chi_L|$) plotted as a function of $\mu$,
displays a diamagnetic background, of order $\mu/\Delta$, with sharp 
paramagnetic peaks of height $\mu/T$ and width $T$.  An exact numerical
 computation
of the expression (21) confirms this qualitative picture (Fig.~1).

Next, we consider temperatures in the range 
$\Delta \ll  T$ \raisebox{-.6ex}{\mbox{$\stackrel{\textstyle <}{ \sim}$}} $
 \Gamma \equiv \hbar v_F/2R$.  For such temperatures
the paramagnetic peaks get smeared and the diamagnetic background, of
order $(k_FR)^2$, cancels with the corresponding paramagnetic term.
The net effect is an oscillating term, of order $(k_FR)^{{3}/{2}}$.  The 
calculation is based on a Poisson summation of the double sum in Eq.~(21)
and on a semi-classical approximation for the unperturbed energies, or
 $\lambda_{mn}$,
which satisfy$^{13}$:
\begin{eqnarray}
\sqrt{\lambda^2_{mn}-m^2} - m \arccos(m/\lambda_{mn})=\pi (n+\frac{3}{4}) 
\end{eqnarray}
Let us outline the calculation of the paramagnetic term\linebreak 
$\chi^{(p)}/|\chi_L|= -
(6\hbar^2/MR^2)\sum m^2(\partial f/\partial \epsilon_{mn})$.  As usual, it is 
simpler to consider first the zero-temperature case and then to use Eq.~(9).
 At
$T=0$, $\partial f/\partial \epsilon = - \delta(\mu-\epsilon)$ and
\begin{eqnarray}
\chi^{(p)}_0(E) =
 12 \, |\chi_L|\sum_{m,n}m^2\delta(\lambda^2_{mn} - \gamma^2) \ , 
\end{eqnarray}
where $\gamma^2=4\mu/\Delta$.  After applying the Poisson summation formula,
$m$ and $n$ go over into continuous variables: $m\rightarrow x$,
 $(n+\frac{3}{4})
\rightarrow y$.  The integral over $y$ is immediate, due to the
 $\delta$-function.
The integral over $x$ is done in the saddle-point approximation, using the 
large parameter $\gamma$ and the continuous version of Eq.~(22).  The
resulting expression for $\chi^{(p)}_0(E)$ is:
\begin{eqnarray}
\chi_o^{(p)}(E) = |\chi_L|\left\{ \frac{3}{4}\gamma^2+\gamma^{{3}/{2}}
\sum_{K_x,K_y} \phi(K_x,K_y) + O(\gamma)\right\} \ ,  
\end{eqnarray}
where the sum runs over $1\leq K_y<\infty$ and $0\leq 2K_x\leq K_y$, and
\begin{eqnarray}
&&\phi(K_x,K_y)=\frac{24}{\sqrt{\pi K_y}} \cos^2\left(\pi\frac{K_x}{K_y}
\right)
\sin^{{3}/{2}} \left( \, \pi\frac{K_x}{K_y}\right)\cdot \mbox{\rule[-.6cm]{0cm}{1cm} } \nonumber \\
&&\hspace{0.50cm} \cdot \cos\left[\,2 \, \gamma \, K_y\sin 
\left(\pi\frac{K_x}{K_y}\right) - \frac{3}{2}\pi K_y + \frac{\pi}{4}\,
\right] \ .  
\end{eqnarray}
A similar treatment of the diamagnetic term results in a term $-3\gamma^2/4$
 plus
oscillating corrections of order $\sqrt{\gamma}$.  Thus, the large
 non-oscillating terms cancel and the net result for $\chi_0(E)$ is 
given by the second term in Eq.~(24).  Finally, using Eq.~(9), (12), we find:
\begin{eqnarray}
\chi=|\chi_L|\,\gamma^{{3}/{2}}\sum_{K_x,K_y} \phi(K_x,K_y)\,{\cal R}
\left(\frac{\,\pi\, K_y T }{\Gamma}\sin\left(\pi\frac{K_x}{K_y}\right)\right) \ .
\end{eqnarray}
Thus, in the temperature range $\Delta \ll T$ \raisebox{-.6ex}{\mbox{$\stackrel{\textstyle <}{ \sim}$}} $ \Gamma$, the susceptibility
$\chi$ is, typically, of order $(k_FR)^{{3}/{2}}$ and can have either sign.
It oscillates, as a function of $\mu$, with a period of order
$\left({\mu\, \Delta}\right)^{1/2} \simeq \Gamma$.  In Fig.~2 we present $\chi$ as a
 function of $\gamma$, as obtained from Eq.~(26) (solid line). 
 For comparison, dots represent the result of a numerical computation,
 based on the exact Eq.~(10), with 
$\epsilon'_i$ and  $\epsilon^{\prime\prime}_i$ given by Eq.~(20).  These
oscillations reflect the structure of the density of states smoothed over
energy intervals $\Delta E$ \raisebox{-.6ex}{\mbox{$\stackrel{\textstyle <}{ \sim}$}} $ \Gamma$, in the same way as the sharp peaks of
the low-temperature regime reflected the exact (discrete) spectrum of the
 system.
Eq.~(26) can be obtained from the corresponding expression of Ref.~4, in the
$B\rightarrow 0$ limit.  The derivation in Ref.~4 was based on a semi-classical
approximation for the density of states$^{8,10}$, with a subsequent
introduction of the magnetic field via the classical action.  In contrast, we have 
started with an exact expression for the linear susceptibility and used (in 
the mesoscopic temperature regime) a semi-classical approximation for the energy
levels of the system.$^{13}$\\

\noindent{\bf 5. Square Geometry}

Here we discuss electrons within a square of size $L$.  The unperturbed
energies are $\epsilon_{nm} = (\pi^2\hbar^2/2ML^2)(n^2+m^2)$.  A state
$|n,m\rangle$ is degenerate with the state $|m,n\rangle$ (there can be,
in addition, accidental degeneracies if a pair $n',m'$ has the same sum of squares
as the pair $n,m$).  

Let us first consider low temperatures, $T\ll\Delta\equiv 2\pi\hbar^2/ML^2$, and 
discuss the paramagnetic peaks due to the second term in Eq.~(10).  The first
order correction $\epsilon'_{nm}$ is due to lifting of the double degeneracies
by the magnetic field. (We do not consider accidental degeneracies, although
the treatment is readily extended to include that case as well.)
The degeneracy is lifted only if $n$ and $m$ have different parity and in that
case
\begin{eqnarray}
\epsilon'_{nm} = - \epsilon'_{mn} = \frac{32}{\pi^2} \frac{\hbar e B}{Mc}
\frac{n^2m^2}{(n^2-m^2)^3} 
\end{eqnarray}
The largest corrections occur when $n=m\pm 1$.  In such cases $\epsilon'_{nm}
\simeq (\hbar e B/Mc)k_FL$ and the height of the corresponding peak in 
susceptibility is $\chi_{\rm max} \simeq |\chi_L|(k_FL)^2\Delta/T$, just as in
the case of the disc.  Note, however, that in the square geometry such large peaks
are much more rate than in a disc.  For a disc large peaks were separated by
a distance of order $\Delta$.  For a square such peaks occur, roughly, for each
pair $(n,n+1)$, i.e. are separated by a distance of order $n\Delta \simeq k_FL\Delta
\simeq \Gamma \equiv \hbar v_F/L$.  The difference between a square and a disc
is clearly seen, if one compares Fig.~3 to Fig.~1.  Except for the large 
paramagnetic peaks in Fig.~3 one can see an oscillating background.  This
background comes form the first term in Eq.~(10).  It will be shown below that
this term can be either para- or diamagnetic and its typical value is of the
order $(k_FL)^{{3}/{2}}$.

In the mesoscopic temperature regime, $\Delta < T < \Gamma$, the
 susceptibility
$\chi$ is described by the semi-classical expression (13).  This case was 
studied in detail in Ref.~6 and particularly in Ref.~4 where expressions for
 the 
factors $d_{\lambda,r}$ and $\langle A^2_\lambda \rangle$ can be found.
The resulting expression for $\chi$ is:
\begin{eqnarray}
\frac{\chi}{|\chi_L|} &=& \frac{8}{5\sqrt{\pi}}(k_FL)^{{3}/{2}}
\sum^\infty_{r=1} \sum_{\matrix{ u_xu_y \cr {\rm odd} } }
\frac{\sin\left(2 r \sqrt{u^2_x+u^2_y}\,k_FL + \pi/4\right)
\mbox{\rule[-.1cm]{0cm}{1cm}}}{\mbox{\rule[-.7cm]{0cm}{1cm}}r^{{1}/{2}}\,
 \left(u^2_x+u^2_y\right)^{{5}/{4}} 
\, u^2_xu^2_y} \cdot 
\nonumber \\
&& \hspace{1cm} \cdot  {\cal R}\left(\frac{2\pi r \sqrt{u^2_x+u^2_y}\, T}
{\Gamma}\right) \ \mbox{\rule[-.1cm]{0cm}{1cm}},   
\end{eqnarray}
where $u_x$ and $u_y$ are positive coprime integers, which label primitive
orbits.  Only odd $u_x$ and $u_y$ enter the sum in Eq.~(28), since otherwise
the area enclosed by an orbit is exactly zero.$^4$

As an example, in Fig.~(4) we present $\chi/|\chi_L|$ as calculated from
 Eq.~(28) (dots).  We chose the same ratio $T/\Delta = 5$ as in Ref.~4. 
 The result is practically
indistinguishable from the numerical one (solid line), based on Eq.~(10).
(Let us note, that in Ref.~4 some disagreement between Eq.~(28) and numerics
was observed).  There are some qualitative differences between the mesoscopic
oscillations in the square geometry (Fig.~4) and those for a circle (Fig.~2):
in the circle oscillations are modulated on an energy scale which exceeds
$\Gamma$ by an order of magnitude.

Comparison with an exact numerical computation demonstrates that expression
(28) is valid down to temperatures $T\approx \Delta$, but fails for $T\ll\Delta$.
Nevertheless, it can be used for estimating the aforementioned background,
 due to the first term in Eq.~(10).  The point is that this term ceases to
 change when temperature is lowered from $T\approx \Delta$ down to $T=0$. 
 To obtain an estimate of expression (28) at low temperatures, we square 
it and average out the fast oscillations.  This gives, for the typical value
 of $\chi^2$ in the background
\begin{eqnarray}
\chi^2_{\rm back} \simeq \chi^2_L\, \frac{32 }{25\pi}\, (k_F L)^3 \sum_{r,u_x,u_y}
\frac{{\cal R}^2\left( 2\pi r \sqrt{u_x^2+u^2_y}\,T\,/\,\Gamma \right)\mbox{\rule[-.2cm]{0cm}{1cm}}}
{\mbox{\rule[-.7cm]{0cm}{1cm}} r\,\left(u_x^2+u^2_y\right)^{{5}/{2}}\, u^4_xu^4_y} 
\end{eqnarray}
The sum over repetitions $r$ is estimated by replacing it with an integral,
with an upper cutoff provided by the function ${\cal R}$.  This gives a
logarithmic factor, so that
\begin{eqnarray}
\left|\chi_{\rm back}\right| \simeq (k_FL)^{{3}/{2}}\sqrt{\ln(k_FL)} \ . 
\end{eqnarray}
Let us mention that a similar logarithmic factor appears in chaotic billiards,
at low temperatures.$^5$  The behaviour of $\chi$ in that case is, of course, 
quite different form the square geometry: the enhancement factor is $k_FL$,
 instead of $(k_FL)^{{3}/{2}}$, and the large paramagnetic peaks disappear.

\pagebreak

\noindent{\bf 6. Harmonic Confinement}

Our last example is a degenerate electron gas confined by a harmonic potential.$^{14-17}$
The problem of orbital magnetism in this case was considered previously by a 
number of authors, and some analytical and numerical results have been obtained
for various temperatures and fields.  Below we shall derive an analytical expression
for the linear susceptibility $\chi$, as a function of temperature.$^{18}$  This 
expression demonstrates the crossover from strong magnetic effects towards the
weak Landau diamagnetism, under increase of temperature.

We consider a slightly anisotropic harmonic potential, \linebreak
\mbox{ $U(x,y)=(M/2)(\Omega_1\,x^2+\Omega_2\,y^2)$},
with $\Omega_1$ close to $\Omega_2$, namely \linebreak
$|\Omega_1-\Omega_2|\equiv {\scriptstyle \Delta} \Omega < \hbar \Omega^2_1/\mu$.  In the absence of a magnetic field the spectrum is given by
 $\epsilon_{nm} = \hbar\Omega_1
(n+\frac{1}{2}) +  \hbar\Omega_2(m+\frac{1}{2})$, where $n,m=1,2,\ldots$.
For the isotropic case, $\Omega_1=\Omega_2=\Omega$, energy levels can be
 labelled by a single integer $\ell=1,2,\ldots$, and the $\ell$-th level is
 $\ell$-fold degenerate.
A small anisotropy ${\scriptstyle \Delta} \Omega$ splits the degenerate levels into 
narrow ``multiplets''.  The width of the $n$-th multiplet is of order
$\ell\hbar{\scriptstyle \Delta} \Omega$, which is $\mu{\scriptstyle \Delta} \Omega/\Omega$ for multiplets near
the energy $\mu$.  The above formulated condition, 
${\scriptstyle \Delta} \Omega<\hbar\Omega^2/\mu$,
is just the requirement for having well defined multiplets.  It is clear on
physical ground, and is verified by the calculation below, that such weak
anisotropy can affect the physical properties of the system only at temperatures,
$T<\mu{\scriptstyle \Delta} \Omega/\Omega$.

When a weak magnetic field is applied, energy levels acquire a second order 
correction
\begin{eqnarray}
\epsilon^{\prime\prime}_{nm} = \frac{1}{2}\hbar\left(\frac{eB}{Mc}\right)^2
\frac{(n+\frac{1}{2})\Omega_1-(m+\frac{1}{2})\Omega_2}{\Omega^2_1-\Omega^2_2} \ .
\end{eqnarray}
Since for any finite anisotropy the first order correction is zero, Eq.~(10) gives:
\begin{eqnarray}
\frac{\chi}{|\chi_L|} = - \frac{24\hbar}{MR^2}\, 
\frac{1}{\Omega^2_1-\Omega^2_2}\, \sum_{n,m}\left[(n+\frac{1}{2})\Omega_1-
(m+\frac{1}{2})\Omega_2\right] f(\epsilon_{nm}) \ ,   
\end{eqnarray}
The effective radius $R$ is determined from $M\Omega^2R^2/2=\mu$, where 
$\Omega=(\Omega_1+\Omega_2)/2\approx\Omega_1$.  Applying to the double sum
in Eq.~(32) the Poisson summation formula, we obtain for the zero-temperature
susceptibility $\chi_0(E)$:
\begin{eqnarray}
\frac{\chi_0(E)}{|\chi_L|} = - \frac{12}{\gamma(1-\lambda^2)} 
\sum^{+\infty}_{l,k=-\infty} (-1)^{l+k} \int^{\gamma/\lambda}_0 dy
\int^{\gamma-\lambda y}_0 dx (x-\lambda y)e^{i2\pi(l y+kx)} 
\end{eqnarray}
where $\lambda = \Omega_2/\Omega_1$ and $\gamma=\mu/\hbar\Omega=k_FR/2$.

Calculating the elementary integrals and making use of Eqs.~(9), (12), 
one can obtain
a final expression for the susceptibility $\chi(T)$.  This expression is 
rather cumbersome and will not be given here.$^9$  For temperatures 
$T \gg \gamma\hbar{\scriptstyle \Delta} \Omega$ and anisotropy 
${\scriptstyle \Delta} \Omega\ll\Omega/\gamma$
it simplifies to:
\begin{eqnarray}
\frac{\chi}{|\chi_L|} = - 1 + 2\gamma^2\sum^\infty_{r=1}
 \cos(2\pi r \gamma)
{\cal R}\left(\frac{\pi^2 r \hspace{0.05cm} T}{\Gamma}\right) 
\end{eqnarray}
where $\Gamma=\hbar v_F/2R=\hbar\Omega/2$.  Except for the  leading 
oscillating
term, of order $\gamma^2$, there are smaller oscillating terms which are 
not written in Eq.~(34).  This equation does not contain anisotropy, which
demonstrates that, for $T$ much larger than the multiplet width 
$\gamma\hbar{\scriptstyle \Delta} \Omega$,
the anisotropy does not come into play (up to exponentially small 
corrections).
For $T \gg \Gamma$, the oscillations in Eq.~(34) are negligible and 
$\chi=\chi_L$.
for $T$ \raisebox{-.6ex}{\mbox{$\stackrel{\textstyle <}{ \sim}$}} $ \Gamma$, 
one can keep only the first term in he sum, which results in 
an oscillating term of order $\gamma^2\simeq (k_FR)^2$.  For $\gamma\hbar
{scriptstyle \Delta}\Omega \ll T \ll \Gamma$, many terms contribute to the
 sum.  The resulting expression
exhibits paramagnetic peaks, of height $(k_FR)^2\Gamma/16T$ and width $T$, 
on a diamagnetic background of order $-(k_FR)^2$.

For $T<\gamma\hbar{\scriptstyle \Delta} \Omega$, Eq.~(34) ceases to be applicable.  An analysis of
the full expression$^9$ for $\chi(T)$ shows that it matches the Eq.~(34)
at $T\approx \gamma \hbar {\scriptstyle \Delta} \Omega$ and that only minor changes in 
$\chi(T)$ occur when $T$ is lowered below $\gamma\hbar{\scriptstyle \Delta} \Omega$.  This 
means that at low temperatures, $T<\gamma\hbar{\scriptstyle \Delta} \Omega$, the width of the 
paramagnetic peaks becomes $\gamma\hbar{\scriptstyle \Delta} \Omega$ and their height is of order
$k_FR\Gamma/\hbar{\scriptstyle \Delta} \Omega$.

This result is a manifestation of a general rule. If there is a cluster of nearly
degenerate levels about some energy $\epsilon_c$ (the width of the cluster $\delta$
is much smaller than the typical level spacing $\Delta$), then for $T>\delta$
the cluster behaves as a single degenerate level: it gives rise to a paramagnetic\
peak of height $g(k_FL)^2\Delta/T$, where $g$ is the number of non-zero
eigenvalues of the matrix $\langle i |\hat{L}_z|j\rangle$ in the subspace of the 
cluster.  For $T<\delta$, the width of the cluster becomes relevant and the
peak saturates at a value of order $g(k_FL)^2\Delta/\delta$.  This is best seen
if $\epsilon'_i$ and $\epsilon^{\prime\prime}_i$ in Eq.~(10) are written explicitly using
the symmetric gauge.  This leads to:
\begin{eqnarray}
\frac{\chi}{|\chi_L|} &=& - \frac{3\pi}{A}\left\{
\sum_i \langle i|\hat{r}^2|i\rangle f(\epsilon_i)\, + \right. \nonumber \\
&+& \left. \frac{1}{M} \sum_{ij}
|\langle i | \hat{L}_z| j \rangle|^2
\int^1_0 dx\cdot f'\left(\epsilon_i-x(\epsilon_i-\epsilon_j)\right)\right\} 
\end{eqnarray}
When $\delta<T<\Delta$ the paramagnetic contribution of the cluster, as given by
the second term in Eq.~(35), is approximately $-(3\pi/AM)f'(\epsilon_c)trL^2_z$,
where the trace is within the subspace of  the cluster.  Estimating the
trace as $g\hbar^2(k_FL)^2$ leads to the expression $g(k_FL)^2\Delta/T$ for the Eq.~
paramagnetic peak.

Let us, finally, mention that when the anisotropy ${\scriptstyle \Delta} \Omega$ approaches the 
value $\Omega/\gamma$ (i.e. neighbouring multiplets start to overlap), the
well pronounced paramagnetic peaks disappear, although the oscillations are
still of order $(k_FR)^2$.\\

\noindent{\bf 7. Conclusions}

We have calculated the linear magnetic susceptibility $\chi(T)$ for several\
two-dimensional integrable systems.  Generally, there are three distinct energy
scales: level spacing $\Delta$, inverse time of flight $\Gamma$ and
the Fermi energy $\epsilon_F$.  

For $T<\Delta$, the susceptibility is sensitive to the  detailed structure of
the energy spectrum as well as the matrix elements of the angular momentum operator.
In particular, degeneracies lead to large paramagnetic peaks of order
$g(k_FL)^2\Delta/T$, $g$ being the level degeneracy.  At such low temperatures
the sample dimensionality is of no importance and similar peaks exist also in 
three-dimensional systems with degenerate levels.$^{19}$  For a pair of nearly 
degenerate levels, when the level separation $\delta\ll\Delta$, the height of the 
corresponding peak saturates at $T\approx \delta$.  The typical value of
$\chi$ between the peaks is also non-universal: for systems with rotational
symmetry it is of order $\chi_L(k_FL)^2$ whereas for a square it is of
order $|\chi_L|(k_FL)^{{3}/{2}}\sqrt{\ln(k_FL)}$ and can be either
para- or diamagnetic.

For the mesoscopic temperatures, $\Delta \ll T$ \raisebox{-.6ex}{\mbox{$\stackrel{\textstyle <}{ \sim}$}} $ \Gamma$, the orbital magnetic 
susceptibility remains large.  In units of $|\chi_L|$, it is of order
$(k_FL)^\alpha$ and oscillates, as a function of the chemical potential $\mu$,
with a period ${\scriptstyle \Delta}\mu\simeq\Gamma$.  For generic integrable systems
$\alpha=\frac{3}{2}$ (compare to $\alpha=1$ for chaotic systems$^3$).  For quasi-one-dimensional
systems (a strip) $\alpha=\frac{1}{2}$ and for a harmonic confining potential 
$\alpha=2$.  Thus, the harmonic potential is a very special case even among
the integrable systems.  The point is that, for the isotropic case and at zero
magnetic field, the two-dimensional harmonic potential exhibits ``accidental''
degeneracies, which are not related to the rotational symmetry (alike
the ``accidental'' degeneracies in the hydrogen atom).

Finally, for $\Gamma \ll T \ll \mu$, all large orbital magnetic effects
disappear and $\chi$ becomes equal to the Landau value $\chi_L$. 
 Thus, for the macroscopic
limit to be achieved, it is not sufficient to have $T \gg \Delta$, as one
 might naively expect.  A much more stringent condition, $T \gg \Gamma$,
 is needed.  We close the paper with the following remarks:
\begin{enumerate}
\item[(i)] Although our calculations have been done for the grand-canonical
ensemble, one can immediately infer about the picture for the canonical ensemble.
For $T\ll\Delta$, the susceptibility can be very sensitive to the exact number
of particles in the system, provided that the last occupied level is degenerate.
If such a level is partially occupied, the system is paramagnetic with $\chi\simeq
c|\chi_L|(k_FL)^2\Delta/T$, where the factor $c$ accounts for the degeneracy and
occupancy of the level.  For a fully occupied level, and in the  presence of
rotational symmetry, the response is diamagnetic, with $\chi\simeq \chi_L
(k_FL)^2$.  (The discussion can be generalised to the case of a cluster of
nearly degenerate levels, as was done above for the grand-canonical ensemble.)
For mesoscopic temperatures, $\Delta \ll T$ \raisebox{-.6ex}{\mbox{$\stackrel{\textstyle <}{ \sim}$}} $ \Gamma$, the standard transition
from the grand-canonical ensemble to the canonical one applies, i.e. $\chi_{\rm can}(N)
= \chi_{\rm gr}(\mu(N)$).  Indeed, the thermal fluctuation $\delta N\simeq
\sqrt{T/\Delta}$ in the number of particles is much smaller than the change
in the number of particles, $({\scriptstyle \Delta}N) \simeq ({\scriptstyle
 \Delta}\mu)(\partial N/\partial \mu)\simeq T/\Delta$,
corresponding to a significant change \linebreak  in $\chi$.
\item[(ii)] So far we have discussed only the orbital magnetic susceptibility.
The Zeeman splitting, due to the electron spin, also contributes.  Within the
linear response its contribution $\chi_s$ is simply added to the orbital part of the
susceptibility.  This is clearly seen form Eq.~(10), if the Zeeman splitting
$\pm \mu_BB$ is included into the first order energy correction $\epsilon'_i$.
This gives $\chi_s=A^{-1}\mu^2_B(\partial N/\partial \mu)$, where
 $N=\sum_if(\epsilon_i)$.
At low temperatures, $T\ll\Delta$, $\chi_s$ exhibits paramagnetic peaks 
$\chi_s\simeq |\chi_L|\,\Delta/T$, which is just the Curie paramagnetism due
 to the last occupied level.  For temperatures $T \gg \Delta$, $\chi_s$ is 
given by the Pauli paramagnetism, $\chi_s=3|\chi_L|$, up to small 
corrections.  Thus, in this case, there is no appreciable mesoscopic effects
 due to the electron spin.
\item[(iii)] The magnetic field in this paper was treated as a given (homogeneous)
external field, $B_{\rm ext}$.  For sufficiently large $\chi$, however, the orbital
magnetic currents flowing in the sample produce a field $B_{\rm curr}$ which is
comparable to $B_{\rm ext}$.  In three-dimensional sample this happens when
 the magnetisation $M$ becomes comparable to $B_{\rm ext}/4\pi$, or 
$|\chi|\simeq 1/4\pi$.  In two dimensions the magnetisation is defined as the
magnetic moment per unit area (rather than per unit volume), so that $\chi$
has units of length.  Also, since the thickness of the sample (height in 
$z$-direction)
is much smaller than its size $L$, $B_{\rm curr}$ differs very much from the 
volume magnetisation (demagnetisation effect). 
 So one have to estimate the field $B_{\rm curr}$ and compare it
 to $B_{\rm ext}$.  The most stringent limitation is set by the requirement,
 that the largest possible value the {\em local} field $B_{\rm curr}$ can
 assume should be much smaller than $B_{\rm ext}$.  This yields$^{9}$ 
$|\chi|k_F\ln(k_F L) \ll 1$ , or 
$|\chi/\chi_L|\ll \ln(k_F L)/k_F r_e$ , where $r_e=e^2/mc^2$ is the classical
 electron radius.  (Though, in some cases, e.g.  in a square geometry,
the condition is less severe, namely $|\chi/\chi_L|\ll L\ln(k_F L)/r_e$.)
 As an example consider a disc geometry, where at paramagnetic peaks $\chi/|\chi_L|\simeq (k_F L)^2 \Delta /T$.  If $\lambda_F\simeq 10^{-6}\ cm$ and 
$k_F L\simeq 100$, then, as long as $T \gg 10^{-3}\Delta$, the self-consistent
 treatment is not necessary.  If the condition is not satisfied, one cannot 
assume anymore a given external field, but should solve the entire problem 
self-consistently.  A self-consistent treatment leads to such possibilities 
as Meissner effect and orbital ferromagnetism, both in bulk samples and rings
.$^{2,3,19-22}$  An interesting possibility appears to exist in the strip 
geometry, considered in Section 3.  Here the linear susceptibility, Eq.~(18),
 is $\chi_0\simeq |\chi_L|\sqrt{k_FL_y}$.  However, the {\em differential}
magnetic susceptibility, $\chi_d$, in a finite field \mbox{ $B$ \raisebox{-.6ex}{\mbox{$\stackrel{\textstyle <}{ \sim}$}} $ mcv_F/eL_y$}, turns 
out to be of order $|\chi_L|(k_FL_y)^{{3}/{2}}$ (see Ref.~12, where a 
strip with harmonic confinement was analysed).$^{23}$  If  \mbox{$|\chi_0| \ll \lambda_F /4\pi \ln(k_F L)$} but $|\chi_d|> \lambda_F /4\pi \ln(k_F L)$,
then one encounters a situation similar to the one which can occur in the 
de Haas-van-Alphen effect and leads to a break up of the sample into magnetic
domains.$^{24}$
\end{enumerate}

This work was supported by the Fund for the Promotion of Research at the Technion.

\pagebreak

\noindent {\bf References}

\begin{enumerate}
\item L.D. Landau and E.M. Lifshitz, Statistical Physics, 3rd Edition,
Part 1, Pergamon Press (1980).
\item V.L. Ginzburg, Uspekhi Fiz.\ Nauk {\bf 48} (1952) 25 [German 
translation in: Fortscht.\ Phys.\ {\bf 1} (1953) 101].
\item B. Shapiro, Physica A200 (1993) 498.
\item K. Richter, D. Ullmo and R.A. Jalabert, (submitted to Physics Reports),
and references therein.
\item O. Agam, J.\ Phys.\ I France {\bf 4} (1994) 697.
\item F. von Oppen, Phys.\ Rev.\ {\bf B50} (1994) 17151.
\item E. Akkermans and B. Shapiro, Europhys.\ Lett.\ {\bf 11} (1990) 467.
\item M.C. Gutzwiller, Chaos in Classical and Quantum Mechanics, Springer (1990).
\item E. Gurevich, Ms. Thesis (Technion).
\item M.V. Berry and M. Tabor, Proc.\ R.\ Soc.\ Lond.\ {\bf A349} (1976) 101.
\item L. Friedman, Phys.\ Rev.\ {\bf 134} (1964) A336.
\item J. Hajdu and B. Shapiro, Europhys.\ Lett.\ {\bf 28} (1994) 61.
\item E.N. Bogachek and G.A. Gogadze, Sov.\ Phys.\ JETP, {\bf 36} (1973) 973.
\item R. Nemeth, Z.\ Phys.\ {\bf B81} (1990) 89.
\item D. Yoshioka and H. Fukuyama, J. Phys.\ Soc.\ Japan {\bf 61} (1992) 2368.
\item Y. Meir, O. Entin-Wohlman and Y. Gefen, Phys.\ Rev.\ {\bf B42} (1990)
8351.
\item M. Azbel, preprint (1996).
\item The analysis can be extended to an arbitrary magnetic field (Ref.~9).
\item See e.g. A.I. Buzdin, O.V. Dolgov and Y.E. Lozovik, Phys.\ Lett.\ {\bf 100A}
(1984) 261.
\item I.O. Kulik, Pisma Zh.\ Eksp.\ Teor.\ Fiz.\ {\bf 11} (1970) 407 
[JETP Lett.\ {\bf 11} (1970) 275]
\item D. Wohlleben, J.\ Less Common Metals {\bf 138} (1988) 11;
D. Wohlleben et al., Phys.\ Rev.\ Lett.\ {\bf 66} (1991) 3191.
\item M.Ya. Azbel, Phys.\ Rev.\ {\bf B48} (1993) 4592.
\item This can be explained in terms of the semi-classical
 picture.  In the strip geometry the leading contribution to $\chi_{osc}$ 
comes from the ``bouncing-ball'' trajectories (at $B=0$ these are 
self-retracing ones).  At zero magnetic field an area enclosed by the
trajectories is exactly zero.  This area increases along with $B$ and becomes 
of order $L_y^2$ at $B \simeq mcv_F/eL_y$, what yields $\chi_d \simeq 
|\chi_L|(k_F L)^{3/2}$ at such a field.
\item I.M. Lifshitz, M.Ya. Azbel, and M.I. Kaganov, Electron Theory of
Metals (Consultant Bureau, New York, 1973).
\end{enumerate}
\pagebreak
{\bf Figures} 

Fig.~1.  Linear magnetic susceptibility in a disc geometry, calculated 
at $T=0.1 \, \Delta$ and normalised to $|\chi_L| (k_F R)^2$.  \\

Fig.~2.  Linear magnetic susceptibility in a disc geometry, calculated
 at $T=5 \, \Delta$ and normalised to $|\chi_L| (k_F R)^{3/2}$.
  Analytic result obtained from Eq.~(26) (solid line) is compared to numerical
one based on Eq.~(35) (dots).\\

Fig.~3.  Linear magnetic susceptibility in a square geometry, calculated
 at $T=0.1 \, \Delta$ and normalised to $|\chi_L| (k_F R)^{3/2}$.  \\

Fig.~4.  Linear magnetic susceptibility in a square geometry, calculated 
at $T=5 \, \Delta$ and  normalised to $|\chi_L|$.  Analytic result obtained 
from  Eq.~(28) (solid line) is compared to numerical one based on Eq.~(35) 
(dots).

\end{document}